\documentstyle[aps,preprint]{revtex}
\begin{document}
\draft
\tighten
\preprint{\vbox{
                \hfill TIT/HEP-400/NP 
 }}
\title{QCD sum rule calculation of the $\pi NN$ coupling-revisited }
\author{Hungchong Kim$^1$ \footnote{E-mail : 
hckim@th.phys.titech.ac.jp},
Su Houng Lee$^{2,3}$\footnote{AvH fellow.}  and
Makoto Oka$^1$ }
\address{$^1$ Department of Physics, Tokyo Institute of Technology, Tokyo 
152-8551, 
Japan \\
$^2$ GSI, Planckstr. 1, D-64291 Darmstadt, Germany \\
$^3$ Department of Physics, Yonsei University, Seoul 120-749, Korea }

\maketitle
\begin{abstract}

In the conventional QCD sum rule calculation for the   
$\pi NN$ coupling beyond the soft-pion limit, there are three
distinct Dirac structures, each of which can be used to
construct separate sum rule. 
We point out subtleties in the previous sum rule results, based on one of the 
Dirac structure,   associated with using either the  
PV or the PS coupling schemes in modeling the  phenomenological side.
We propose a sum rule coming from a different Dirac 
structure, which is independent of the coupling schemes used and 
has less uncertainty in the OPE.  The obtained value is 
$g_{\pi N}=9.76 \pm 2.04$ where the uncertainty mainly comes
from the quark-gluon mixed condensate.

\end{abstract}

Since first introduced by Shifman, Vainshtein and Zakharov~\cite{SVZ},
QCD sum rule has been widely used to study the hadron properties such as
masses or couplings of baryons~\cite{qsr}. 
QCD sum rule is a framework which connects the physical parameters with
QCD parameters.  In this framework, a correlation function is 
introduced in terms of  interpolating fields constructed from quark and
gluon fields.
Then, the correlation function, 
on the one hand,  is calculated by Wilson's
operator product expansion (OPE) and, on the other hand, its  phenomenological
``ansatz'' is constructed.  A physical quantity of
interest is extracted by matching the two descriptions in the
deep Euclidean region ($q^2 = - \infty$) making use of the dispersion relation.
The extracted parameter  therefore should be independent of  the possible
ansatz in order to be physically meaningful.

One important quantity to be determined in hadron physics is the pion-nucleon
coupling constant, $g_{\pi N}$.  Empirically its value is known to be around 
13.4 but it is of interest
to determine the coupling from QCD.
QCD sum rule can be used for such purpose and indeed there are such 
calculations of $g_{\pi N}$~\cite{qsr,hat,krippa1,krippa2}.  
Reinders, Rubinstein and Yazaki~\cite{qsr} calculated
$g_{\pi N}$ by retaining only the first nonperturbative term in 
OPE.  Later Shiomi and Hatsuda (SH)~\cite{hat} improved
the calculation by including higher order terms in OPE. SH considered
the two-point correlation function for the nucleon interpolating field
$J_N$,
\begin{eqnarray}
\Pi (q, p) = i \int d^4 x e^{i q \cdot x} \langle 0 | T[J_N (x) 
{\bar J}_N (0)]| \pi (p) \rangle \ ,
\label{two}
\end{eqnarray}
and evaluated the OPE in the soft-pion limit ($p_\mu \rightarrow 0$).

More recently, Birse and Krippa (BK)~\cite{krippa1,krippa2}  pointed out that
the use of the soft-pion limit does not constitute an independent sum
rule from the nucleon sum rule because in the limit the correlation function
is just a chiral rotation of the nucleon correlation function,
\begin{eqnarray}
\Pi (q) = i \int d^4 x e^{i q \cdot x} \langle 0 | T[J_N (x) 
{\bar J}_N (0)] | 0 \rangle \ .
\end{eqnarray}
Therefore, BK considered the sum rule beyond the soft-pion limit.
However, in their approach, the OPE contains some condensates
which are not precisely known. This causes large uncertainty in their result,
$g_{\pi N} = 12 \pm 5$.

In this letter, we construct an alternative sum rule for
$g_{\pi N}$ which contains less uncertainty. 
We start by noting that
the previous calculations have  coupling scheme dependence,  
namely the dependence on either the 
pseudoscalar or the  pseudovector type of interaction in the construction
of the phenomenological side.
We stress the importance of doing the sum rule calculation independent
of the coupling scheme.  
To see the origin of the problem, consider saturating 
Eq.(\ref{two}) with a nucleon intermediate state.  Then we have,
\begin{eqnarray}
\Pi(q,p) \sim \lambda_N { \not\!q+m \over 
q^2-m^2 } \langle N(q)| {\bar J}_N | \pi(p) \rangle.
\end{eqnarray}
The coupling, $g_{\pi N}$, is defined to be the residue of the last term.
\begin{eqnarray}
\langle N(q)| {\bar J}_N | \pi(p) \rangle\sim
{ \not\!q- \not\!p +m \over  (q-p)^2-m^2 } \lambda_N  
g_{\pi N} + non~pole~term
\end{eqnarray}
How to treat the {\it non pole term} will depend on the coupling scheme
used and as we will see, influence the subsequent sum rule analysis for 
$g_{\pi N}$

To be more specific, the pion-nucleon coupling has a unique definition as
a matrix element of a pion field sandwiched by the nucleon states.
This matrix element can appear in Eq.(\ref{two}) by, with the use 
of LSZ reduction, 
moving the pion state into the time-ordered product and inserting
on-shell nucleon intermediate state.  This corresponds to the
residue of the double pole while the rest contains the single nucleon pole
which has dependence on the coupling scheme and therefore needs to be
treated carefully. 
The calculation beyond the soft-pion limit is natural  in this consideration. 
When starting from Eq.~(\ref{two}), 
to leading order in the  pion momentum,
there are three distinct Dirac structures which can in principle
be used to calculate $g_{\pi N}$. 
We discuss them separately and propose
a new sum rule to determine $g_{\pi N}$.

Regarding the existing sum rule calculations of 
$g_{\pi N}$~\cite{qsr,hat,krippa1,krippa2},
one question is whether the results have any dependence on
the $\pi NN$ coupling scheme adopted.  This issue, within our
knowledge, has not been properly considered but it can provide
important aspect of how one can proceed in the construction of the sum rule. 
This question is not specific to $\pi NN$ coupling and it can generally
apply to any sum rule calculation involving the pion. 
 
In QCD sum rule, the phenomenological side
of the correlator is usually constructed by making an ``ansatz'' 
based on some effective models.  
In the case of 
the two-point correlation function in Eq.~(\ref{two}), one often 
uses the pseudoscalar interaction Lagrangian,
\begin{eqnarray}
{\cal L}_{ps} = g_{\pi N} {\bar \psi} i \gamma_5 {\mbox {\boldmath $\tau$}}
\cdot {\mbox {\boldmath $\pi$}} \psi\ ,
\label{ps}
\end{eqnarray}
in constructing the phenomenological side. 
However,  the pseudovector interaction Lagrangian,
\begin{eqnarray}
{\cal L}_{pv} = {g_{\pi N} \over 2m}  {\bar \psi} \gamma_5 \gamma_\mu
 {\mbox {\boldmath $\tau$}} \cdot \partial^\mu {\mbox {\boldmath $\pi$}} 
\psi\ ,
\label{pv}
\end{eqnarray}
can also be used for constructing the phenomenological side.
At present, we do not know which Lagrangian is more reliable in 
describing the pion interaction with nucleons.
The two descriptions are equivalent when the participating
nucleons are on-shell
but they are usually not when the nucleons are 
off-shell.
Since on-shell properties of a
particle are extracted by matching the OPE with the
phenomenological side in the deep Euclidean region, the pion
coupling constant determined from QCD sum rules may have the coupling
scheme dependence. 
Let us employ the two descriptions in the phenomenological side of the
sum rule and see if they lead to any difference.

Using the interaction Lagrangians in the
two-point correlation function, Eq.~(\ref{two}), we 
find that there are four distinct Dirac structures,
(1) $\gamma_5$, (2) $\gamma_5 \sigma_{\mu \nu}$, 
(3) $\gamma_5$ $\not\!p$,
(4) $\gamma_5$ $\not\!q$.   
The fourth structure appears only in the pseudovector case with
the residue factor, $p^2 - 2 p \cdot q$ and this structure is definitely
scheme-dependent. 
In principle, any of these structures can be used to calculate 
the coupling. 
Shiomi and Hatsuda (SH)~\cite{hat} considered the $\gamma_5$ structure in the
soft-pion limit.  

When the 
pseudoscalar interaction Lagrangian
is used, the $\gamma_5$ structure without taking the
soft-pion limit takes the form
\begin{eqnarray}
-{i g_{\pi N} \lambda_N^2 \over (q-p)^2 - m^2 + i \epsilon }
+{i g_{\pi N} \lambda_N^2  p \cdot q \over 
(q^2 - m^2 +i \epsilon)[(q-p)^2 - m^2 + i \epsilon] } + \cdot \cdot \cdot
\end{eqnarray}
where $\lambda_N$ is coupling of the nucleon interpolating field, $J_N$,
to the physical nucleon, $m$ is the nucleon mass.  The dots indicate
the contribution from continuum whose explicit forms are not necessary 
for our discussion. In the soft-pion limit,
the double pole structure disappears, leaving only the simple pole
which was used by SH.

The corresponding expression in the pseudovector coupling scheme is
\begin{eqnarray}
{i g_{\pi N} \lambda_N^2  p^2/2 \over 
(q^2 - m^2 +i \epsilon)[(q-p)^2 - m^2 + i \epsilon]} + \cdot \cdot \cdot\  .
\end{eqnarray}
A crucial difference from above is that now the correlator contains 
the double pole only.  More importantly, in the soft-pion limit, the
correlator is zero and one can not extract any information about the coupling.
Clearly, the method by SH is scheme-dependent. 
Another problem is that since SH considered
the simple pole, the additional simple pole
coming from $N \rightarrow N^*$ transition~\cite{ioffe1}
 can not be separated from
the nucleon simple pole.

What will then be the reliable procedure  
using this $\gamma_5$ structure ?  The common pole in 
both coupling schemes is the double pole.  
Consider expanding the denominator of the double pole
in terms of the pion momentum and keeping only the leading term.  The
resulting
term can also be obtained by taking the following kinematical condition,
\begin{eqnarray}
p^2 = 2 p\cdot q\ .
\label{cond}
\end{eqnarray}
Note that this kinematical condition is also the consequence 
of the on-shell condition
for the participating nucleons, $q^2 =m^2$ and $(q-p)^2 = m^2$,
the condition in which  the physical 
$\pi NN$ coupling should be  defined. 
With this condition, the residues of the double poles are the 
same and in principle
the double pole  
can provide the results independent of the coupling schemes.
A nice aspect of this kinematical condition is that
the scheme-dependent Dirac structure, namely $\gamma_5$$\not\!q$, 
which appears only in the pseudovector
scheme, disappears 
and there are only three distinct Dirac structures common to 
both schemes.

Now we consider $\gamma_5 \not\!p$ structure as studied by Birse and
Krippa (BK).   Using the pseudoscalar interaction Lagrangion,
the phenomenological side of this structure takes the form
\begin{eqnarray}
-{i g_{\pi N} \lambda_N^2  m  \over 
(q^2 - m^2 +i \epsilon)[(q-p)^2 - m^2 + i \epsilon]} + \cdot \cdot \cdot\ . 
\end{eqnarray}
BK carried out the calculation using this phenomenological form 
with taking $p_\mu = 0$ in the denominator
and their sum rule formula can
be succinctly written as 
\begin{eqnarray}
g_{\pi N} + A M^2 = f(M)\ .
\label{ksum}
\end{eqnarray}
$f(M)$ is a  function of the Borel mass $M$. The simple pole contribution
associated with the transition $N \rightarrow N^*$ is contained in the
unknown constant $A$.  To estimate the contribution from $A$,
BK first neglected $A$  and determined $g_{\pi N}$. In their second method,
they applied the operator $1 - M^2 \partial / \partial M^2$ to eliminate
the uncertainty associated with $A$.  This second method is 
equivalent~\cite{ioffe3}
to constructing the sum rule for  $(q^2 - m^2) \Pi (q,p) $, which was
proposed by Jin~\cite{jin}. 
Anyway, the two prescriptions of BK yield the results differed
by less than 5 \% and therefore the simple pole contribution from
the continuum is claimed to be negligible.

However, the statement about smallness of $A$ is actually scheme-dependent.  
If the pseudovector interaction Lagrangian is used, the
phenomenological side for the $\gamma_5$$\not\!p$ structure
takes the form 
\begin{eqnarray}
-{i g_{\pi N} \lambda_N^2 \over 2 m} {1 \over (q-p)^2 - m^2 + i \epsilon }
-{i g_{\pi N} \lambda_N^2  m \over 
(q^2 - m^2 +i \epsilon)[(q-p)^2 - m^2 + i \epsilon] } + \cdot \cdot \cdot\ .
\end{eqnarray}
Note that the double pole is the same as above but
we have an additional simple pole of the nucleon which in principle
could change the results. 
If this were used in BK's sum rule, then instead of Eq.~(\ref{ksum}),
BK  would have obtained  
\begin{eqnarray}
g_{\pi N} \left (1 - {M^2 \over 2 m^2} \right ) + A M^2 = f(M)\ .
\label{ksum2}
\end{eqnarray}
Since $M \sim m \sim 1$ GeV, the value of the  new term is about 0.5. 
If we neglect $A$ as BK did in their first method, then this sum rule 
could give the result twice of
what BK obtained.  On the other hand, the second method of BK leads to 
the same result of  BK, indicating that the contribution from $A$
could be large.

One more Dirac structure, $\gamma_5 \sigma_{\mu\nu}$, contains the
double pole only, independent of the coupling scheme.
The common phenomenological side for this structure is given by
\begin{eqnarray}
-{ g_{\pi N} \lambda_N^2  p^\mu q^\nu  \over 
(q^2 - m^2 +i \epsilon)[(q-p)^2 - m^2 + i \epsilon]} + \cdot \cdot \cdot\ . 
\end{eqnarray}
This structure is zero in the soft-pion limit, but beyond the
soft-pion limit as BK did, this can provide additional sum rule 
for $g_{\pi N}$ independent from the BK sum rule. Since there is no simple 
nucleon pole in this case, the simple pole structure comes only from 
$N \rightarrow  N^*$.  
With this in mind,  we will construct the sum rule for this structure.  
As we will see, this structure has also less uncertainty in the OPE.

Now, we construct QCD sum rule for $g_{\pi N}$ by
considering $\gamma_5 \sigma^{\mu\nu} p_\mu q_\nu$ structure.
Specifically, we consider the correlation function
\begin{eqnarray}
\Pi (q,p) = i \int d^4 x e^{i q \cdot x} \langle 0 | T[J_p (x) 
{\bar J}_n (0)]| \pi^+ (p) \rangle\ .
\label{two2}
\end{eqnarray}
Here $J_p$ is the proton interpolating field of Ioffe~\cite{ioffe2},
\begin{eqnarray}
J_p = \epsilon_{abc} [ u_a^T C \gamma_\mu u_b ] \gamma_5 \gamma^\mu d_c
\end{eqnarray}
and the neutron interpolating field $J_n$ is obtained by replacing
$(u,d) \rightarrow (d,u)$.  
In the OPE,  we will only keep the diquark component of the 
pion wave function and use the vacuum saturation hypothesis
to factor out 
higher dimensional operators in terms of the pion wave function and the 
vacuum expectation value.
This will be more or less similar to BK, with 
 some distinctions as we will see.

The calculation of the correlator, Eq.~(\ref{two2}), in the coordinate
space contains the following quark-antiquark  component of the pion wave 
function, 
\begin{eqnarray}
D^{\alpha\beta}_{a a'} \equiv 
\langle 0 | u^\alpha_a (x) {\bar d}^\beta_{a'} (0) | \pi^+ (p) \rangle\ .
\end{eqnarray}
Here, $\alpha$ and $\beta$ are Dirac indices, $a$ and $a'$ are
color indices. 
The other quarks are contracted to form quark propagators.
This matrix element can be written in terms of three
Dirac structures,
\begin{eqnarray}
D^{\alpha\beta}_{a a'} = && {\delta_{a a'} \over 12} 
(\gamma^\mu \gamma_5)^{\alpha \beta}
\langle 0 |  
{\bar d} (0) \gamma_\mu \gamma_5  u (x) | \pi^+ (p) \rangle\ 
+ {\delta_{a a'} \over 12 } (i \gamma_5)^{\alpha \beta} 
\langle 0 |  
{\bar d}(0) i \gamma_5  u (x) | \pi^+ (p) \rangle\nonumber \\ 
&& - {\delta_{a a'} \over 24} (\gamma_5 \sigma^{\mu\nu})^{\alpha\beta}
\langle 0 |  
{\bar d}(0) \gamma_5 \sigma_{\mu\nu}  u (x) | \pi^+ (p) \rangle\ .
\label{dd}
\end{eqnarray}
These matrix elements can be written in terms of pion wave 
functions~\cite{bely}. 
Since we are doing the calculation up to the first order of $p_\mu q_\nu$,
we need only the overall normalization of the wave functions. 
In fact, to leading order in the pion momentum, 
the first and third matrix elements are given as~\cite{bely},
\begin{eqnarray}
\langle 0 | {\bar d} (0) \gamma_\mu \gamma_5  u (x) | \pi^+ (p) \rangle\ 
&=& i \sqrt{2} f_\pi p_\mu + O(x^2) \label{d1} \\
\langle 0 |{\bar d}(0) \gamma_5 \sigma_{\mu\nu}  u (x) | \pi^+ (p) \rangle\ 
&=&i \sqrt{2} (p_\mu x_\nu - p_\nu x_\mu) 
{f_\pi m_\pi^2 \over 6 (m_u + m_d)}\ .
\label{d2}
\end{eqnarray}
In Eq.~(\ref{d1}), $O(x^2)$ contains the twist 4 term whose contribution
was denoted by $\delta^2$ in Ref.~\cite{krippa1,krippa2}.  This term
does not contribute to our sum rule up to the dimension we consider below.
Note that in Eq.~(\ref{d2}) the factor $f_\pi m_\pi^2 / (m_u + m_d)$ can
be written as $-\langle {\bar q} q \rangle / f_\pi$ by making use of
the Gell-Mann$-$Oakes$-$Renner relation.  
Although the operator seems gauge 
dependent, it is understood that the fixed point gauge is used throughout.  
It is then interesting to note that 
the left hand side of  Eq.~(\ref{d2}) can also be expanded in $x$ such 
that the matrix element that contributes is effectively one with higher
dimension,
\begin{eqnarray}
\langle 0 |{\bar d}(0) \gamma_5 \sigma_{\mu\nu}  D_\alpha  
u (0) | \pi^+ (p) \rangle
 =i \sqrt{2} (p_\mu g_{\alpha \nu} - p_\nu g_{\alpha \mu}) 
{f_\pi m_\pi^2 \over 6 (m_u + m_d)}\ .
\label{su}
\end{eqnarray}

The first term in Eq.~(\ref{dd}) is the one considered in BK's sum rule for
the $\gamma_5$$\not\!p$ structure.
However,  
the other two matrix elements can contribute 
to the $\gamma_5$$\not\!p$ structure as the calculation is being done
beyond the soft-pion limit.
Indeed, up to the highest dimension 7 that BK 
considered,
the third matrix element contributes to the $\gamma_5$$\not\!p$
structure. Note that the chirality of the third term is
different from the first term.
Because of the chirality difference, the chiral odd condensate
contributes to $\gamma_5$$\not\!p$ when the third matrix element
is taken for $D^{\alpha \beta}_{aa'}$.  That is, if we include the 
third term of Eq.~(\ref{dd}), because it is effectively dimension 4
[ Eq.~(\ref{su})],
one dimension higher than Eq.~(\ref{d1})~\footnote{
The physical dimension of Eq.~(\ref{d1}) 
is actually 2.  Ordinarily, such operator in vacuum is of 
dimension 3, which is the way  that BK counted the dimension of
Eq.~(\ref{d1}).  In this way of counting, Eq.~(\ref{su}) is
of dimension 4.},
we could have an additional odd dimensional
operator, $\langle \bar{q} q \rangle$, to form a dimension 7 
operator. 
It only changes the highest dimensional operator 
in the BK's sum rule [Eq.(12) of Ref.~\cite{krippa1}] 
but its contribution is much larger than
their highest operator because it contains the quark operator. 

In our sum rule for the $\gamma_5 \sigma^{\mu\nu} p_\mu q_\nu$ structure,
the second matrix element in Eq.~(\ref{dd}) does not contribute 
up to dimension 7. 
It is straightforward to calculate the OPE and up to dimension 7 and we
obtain
\begin{eqnarray}
- \sqrt{2} 
\langle {\bar q} q \rangle \left [ {{\rm ln} (-q^2) \over 12 \pi^2 f_\pi}
+ {4 \over 3 } {f_\pi \over q^2}  -  
\left \langle {\alpha_s \over \pi} {\cal G}^2 
\right \rangle {1 \over 216 f_\pi q^4} 
+{m_0^2 f_\pi \over 6 q^4}
\right ]\ .
\end{eqnarray} 
In obtaining  the first and third terms, we have used Eq.~(\ref{d2}) while 
the second is  obtained by taking the first 
term in Eq.~(\ref{dd})
for the matrix element $D^{\alpha\beta}_{aa'}$  and replacing one
propagator with the quark condensate. 
The fourth term is also obtained by taking the first term in
Eq.~(\ref{dd}) but in this case other quarks are used to form
the dimension five mixed condensate, 
$\langle {\bar q} g_s \sigma \cdot {\cal G} q \rangle $,
which is usually parameterized in terms of the quark condensate,
$m_0^2 \langle {\bar q} q \rangle $\ .  
The value of $m_0^2$ is not well-known.   From Ref.~\cite{mix}, we take 
the range,  
$0.6 \le m_0^2 \le 1.4$ GeV$^2$ and 
see the sensitivity of our results. 
One interesting aspect
of our OPE expression is that the quark condensate, which could be the 
one important source of uncertainty  in the final result, 
is just an overall factor.

By matching the OPE expression with its corresponding phenomenological 
side after taking Borel transformation, we  obtain
\begin{eqnarray}
g_{\pi N} \lambda_N^2 \left [ B + {1 \over M^2} \right ] e^{-m^2/M^2}
=
- {\langle {\bar q} q \rangle \over f_\pi}
\left [ {M^2 E_0 (x_\pi) \over 12 \pi^2 }
+{4 \over 3 } f^2_\pi  + \left \langle {\alpha_s \over \pi} {\cal G}^2
\right \rangle {1 \over 216  M^2} 
-{m_0^2 f^2_\pi \over 6 M^2}
\right ]\ .
\label{newsum}
\end{eqnarray}
Here $x_\pi = S_{\pi N}/M^2$ with $S_{\pi N}$ being the continuum
threshold, and 
$E_n (x) = 1 -(1+x+ \cdot \cdot \cdot + x^n/n!)~ e^{-x}$ .
The unknown parameter $B$ comes from the simple pole of $N \rightarrow N^*$. 
To eliminate the parameter $\lambda_N$ in the LHS, we use the 
chiral-odd nucleon sum rule which is given by
\begin{eqnarray}
m \lambda_N^2  e^{-m^2/M^2} = - {M^4 E_1 (x_N) \over 4 \pi^2} 
\langle {\bar q} q \rangle  + {1 \over 24 } \langle {\bar q} q \rangle
\left \langle {\alpha_s \over \pi} {\cal G}^2
\right \rangle
\label{nuclo}
\end{eqnarray}
where $x_N=S_N/M^2$ with $S_N$ being the continuum threshold for 
the nucleon sum rule.  
As BK did, we take the ratio of Eqs.~(\ref{newsum}) and (\ref{nuclo}) to
obtain
\begin{eqnarray}
g_{\pi N} (B M^2 + 1)    \left ({f_\pi \over m}\right ) &\equiv& a  + b M^2 
\nonumber \\
& =&
{M^4 E_0 (x_\pi)/3 + 16 \pi^2 f_\pi^2 M^2/3 + 
\pi^2 \left \langle {\alpha_s \over \pi} {\cal G}^2 \right \rangle/54 
-2\pi^2 m_0^2 f_\pi^2 /3 
\over M^4 E_1(x_N) - 
\pi^2 \left \langle {\alpha_s \over \pi} {\cal G}^2 \right \rangle/6 }\ .
\label{oursum}
\end{eqnarray}
It is now clear why our sum rule has advantages over BK's sum rule.
In BK's sum rule, there are two main sources for the error. First
one is the uncertainty in the twist-4 term, $\delta^2$ in their notation,
which yields
$\pm 2$ error in their
final result for $g_{\pi N}$.  
The additional $\pm 2$ error comes from the uncertainty in
the value of the quark condensate. These large errors are due to
the fact that these two
terms are the main terms in their OPE. 
No such
uncertainties enter in our sum rule. 
Instead, we have different sources for the error, $m_0^2$ and 
the gluon condensate.
Note that the two sources  are the highest
dimensional term in our OPE. Therefore, they should be  suppressed in 
the Borel window  chosen. However, as $m_0^2$ is very uncertain,
the result has  some dependence on this parameter.
The error due to $m_0^2$ will be investigated in our numerical calculation.
The additional error due to the gluon condensate,
which we take $\left \langle {\alpha_s \over \pi} 
{\cal G}^2 \right \rangle =(0.33~ {\rm GeV})^4$ in our numerical
calculation, is very small because firstly
its contribution to the sum rule is very small and secondly its
variation is canceled in the ratio, Eq.~(\ref{oursum}).

To extract the value of $g_{\pi N}$, we take different approach from BK.
Normally, the unknown constant $B$ is eliminated by applying the
operator, $1 - M^2 \partial /\partial M^2$, to both sides of the
sum rule.  It is fine mathematically  but, as the OPE does not represent
the total strength of the correlator, this operation
for extracting some physical parameter is rather dangerous. 
For example, when this operation is applied to Eq.~(\ref{nuclo}) after
multiplying $e^{m^2/M^2}$, the Borel curve for $m \lambda_N^2$ is quite
different from the one obtained  before the operation, depending on
the Borel mass but more or less about 20 \% difference. 
Furthermore, the curve for $g_{\pi N}$, 
if obtained from this operation on Eq.~(\ref{oursum}), is
not stable with respect to $M^2$. 
Such $g_{\pi N}$
is just a decreasing function of $M^2$, showing no stability .
Nevertheless, $g_{\pi N}$ obtained by following the two methods of BK around
$M^2 \sim 1 $ GeV$^2$ is about 9.2 or 11.8.

In our analysis, we directly fit the RHS of Eq.~(\ref{oursum}) with 
the function $ a  + b M^2 $ within a reasonable Borel window  and 
determine the coefficients $a $ and $b$.
To check the reliability of the fitting process, we also calculate
$\chi^2 = \sum_{i=1}^N [RHS(M_i) - LHS(M_i)]^2 / N $.   
The nucleon continuum $S_N$ is set to be 2.07 GeV$^2$ which
corresponds to the mass squared of the Roper resonance.  
By restricting the continuum contribution
to be around 50 \% of the first term in the nucleon sum rule of
Eq.~(\ref{nuclo}),
the maximum Borel mass is found to be
around $M^2_{max} \sim 1.24$ GeV$^2$. 
The minimum
Borel mass is determined by requiring the highest power correction in
the pion sum rule of Eq~(\ref{newsum})
to be less than 10 \% of the total OPE. 
Such chosen window
provides the common region of the two sum rules. 
For example, using $m_0^2 = 1$ GeV$^2$, this condition leads to $M^2_{min}
\sim 0.82$ GeV$^2$. 
We take $S_{\pi N}$ to be equal to $S_N$ in our analysis. But
the sensitivity to $S_{\pi N}$  
is very small.  For $S_{\pi N} -S_N = \pm 0.5 $ GeV$^2$,
$g_{\pi N}$ changes only by $\pm 0.3$.  
Within the Borel window, the parameters, $a$ and $b$, are
determined by the best fitted method.  Then, the coupling, $g_{\pi N}$,
is obtained via $g_{\pi N}=m a /f_\pi$ using the physical values,
$m = 0.94 $ GeV and $f_\pi =0.093 $ GeV.

The results are listed in Table~\ref{tab} for various values of $m_0^2$.
First note that the contribution from the unknown
single pole, $N \rightarrow N^*$,  represented by $b$ is relatively small. 
At $m_0^2 = 0.6 $ GeV$^2$, this contribution is almost negligible.
Our result is $g_{\pi N} = 9.76 \pm 2.04$ which is rather smaller
than its empirical value of 13.4. 
It is however possible to obtain  a larger value by reducing $m_0^2$,
by relaxing the 10 \% restriction for $M_{min}^2$ or by restricting the
continuum further. 
Any of these moves the Borel window to smaller Borel masses so that the
Borel curve contains more contribution from higher nonperturbative operators.  
So our result should be
interpreted within the standard prescription of the conventional QCD sum rule.

In summary, we have constructed QCD sum rule to determine  $\pi NN$ 
coupling constant.  We noted that the previous calculations have
the coupling scheme dependence and the determined value of the coupling 
suffers from large uncertainties in QCD parameters.  
We have discussed the issue related to the coupling scheme dependence and 
stressed that
one should 
look into the double pole  beyond the soft-pion limit
in the QCD sum rule study of the  $\pi NN$ coupling. 
In this work, we have proposed
to look at the Dirac structure,  $\gamma_5 \sigma^{\mu\nu}$, beyond the
soft-pion limit. This structure is, first of all, independent
of the effective models employed in the phenomenological side and further
provides the  $\pi NN$ coupling with less
uncertainties from QCD parameters. 
It would also be interesting to purse similar work with a 
different nucleon interpolating field.

\acknowledgments
This work is supported in part by the Grant-in-Aid for JSPS fellow, and  
the Grant-in-Aid for scientific
research (C) (2) 08640356 
of  the Ministry of Education, Science, Sports and Culture of Japan.
The work of  H. Kim is supported by Research Fellowships of
the Japan Society for the Promotion of Science.
The work of S. H. Lee is supported by  KOSEF through grant no. 971-0204-017-2 
and 976-0200-002-2 and  by the 
Korean Ministry of Education through grant no. 98-015-D00061.

\begin{table}
\caption{ The best fitted parameters for $a$,  $b$ and $g_{\pi N}$
at given values of $m_0^2$. For a given value of $m_0^2$, $M_{min}^2$
is chosen such a way that the highest dimensional operators
contribute less than 10 \% of the total OPE. $M_{max}^2$ is chosen so that
the continuum contribution is less than 50 \% of the first term of the
OPE in the nucleon sum rule.  Such chosen $M_{max}^2$ is
1.24 GeV$^2$. Then, $a$ and $b$ are
determined from the best fitting method within the window.
$g_{\pi N}$ is obtained by using physical values
for nucleon mass and the pion decay constant, $m = 0.94$ GeV and
$f_\pi =0.093$ GeV.}

\begin{center}
\begin{tabular}{cccccc}
 $m_0^2$ (GeV$^2$)& $M_{min}^2$ (GeV$^2$) & $a$ & $b$ (GeV$^{-2}$)& 
$\chi^2$ & $g_{\pi N} = m a /f_\pi$ \\
\hline\hline
0.6 & 0.55 & 1.17 & 0.05& $6.3\times 10^{-4}$ & 11.8 \\
0.8 & 0.69 & 1.01 & 0.18& $1\times 10^{-4}$ & 10.22 \\
1.0 & 0.82 & 0.9 & 0.27& $1.7\times 10^{-5}$ & 9.14 \\
1.2 & 0.94 & 0.83 & 0.32& $2\times 10^{-6}$ & 8.35 \\
1.4 & 1.05 & 0.76 & 0.36& $2\times 10^{-7}$ & 7.72  \\
\end{tabular}
\end{center}
\label{tab}

\end{table}
\end{document}